\documentclass[useAMS,usenatbib]{mn2e}
\usepackage{graphicx,amsmath,amssymb}

\newcommand{\eg}{{\it e.g.}}
\newcommand{\ie}{{\it i.e.}}

\newcommand{\propmin}{{\em min}}
\newcommand{\propmed}{{\em med}}
\newcommand{\propmax}{{\em max}}

\newcommand{\citeeq}[1]{Eq.~(\ref{#1})}

\newcommand{\citeeqss}[2]{Eqs.~(\ref{#1})~and~(\ref{#2})}

\newcommand{\citesec}[1]{Sec.~\ref{#1}}

\newcommand{\citetab}[1]{Table~\ref{#1}}
\newcommand{\citefig}[1]{Fig.~\ref{#1}}

\newcommand{\ben}{\begin{eqnarray}}
\newcommand{\een}{\end{eqnarray}}
\newcommand{\be}{\begin{equation}}
\newcommand{\ee}{\end{equation}}
\newcommand{\nn}{\nonumber}

\newcommand{\gev}{\mbox{${\rm GeV}$}}
\newcommand{\percc}{\mbox{${\rm cm^{-3}}$}}

\newcommand{\dphiusr}{\mbox{${\rm cm^{-2}~s^{-1}~GeV^{-1}~sr^{-1}}$}}

\newcommand{\greenff}[2]{\mbox{${\cal G}(#1 \leftarrow #2)$}}

\graphicspath{{./Figs/}}

\title[]{Impact of the spectral hardening of TeV cosmic rays on the prediction
  of the secondary positron flux}

\author[J. Lavalle]
{J. Lavalle$^{1}$\thanks{E-mail: julien.lavalle@uam.es}\thanks{Multidark fellow}
\\
Departamento de F\'isica Te\'orica UAM \& 
Instituto de F\'isica Te\'orica UAM/CSIC\\ 
Universidad Aut\'onoma de Madrid, Cantoblanco, 28049 Madrid --- Spain
}

\begin{document}

\date{\today}

\pagerange{\pageref{firstpage}--\pageref{lastpage}} \pubyear{2002}

\maketitle

\label{firstpage}

\begin{abstract}
The rise in the cosmic-ray positron fraction measured by the PAMELA satellite 
is likely due to the presence of astrophysical sources of positrons, 
\eg~pulsars, on the kpc scale around the Earth. Nevertheless, assessing
the properties of these sources from the positron data requires a good 
knowledge of the secondary positron component generated by the interaction of
cosmic rays with the interstellar gas. In this paper, we investigate the impact 
of the spectral hardening in the cosmic-ray proton and helium fluxes recently 
reported by the ATIC2 and CREAM balloon experiments, on the predictions of the
secondary positron flux. We show that the effect is not negligible, leading
to an increase of the secondary positron flux by up to $\sim$60\% above 
$\sim$100 GeV. We provide fitting formulae that allow a straightforward 
utilization of our results, which can help in deriving constraints on one's 
favorite primary positron source, \eg~pulsars or dark matter.
\end{abstract}


\begin{flushleft}
  Preprint IFT-UAM/CSIC-10-77
\end{flushleft}

\section{Introduction}
\label{sec:intro}

The increase in the positron fraction above a few GeV reported by the PAMELA
collaboration~\citep{2009Natur.458..607A} has triggered a lot of interpretation 
attempts, but it is now likely that it is due to positrons originating from 
conventional astrophysical sources, like pulsars 
\citep[\eg][]{1970Natur.227..465S,1970ApJ...162L.181S,1987ICRC....2...92H,1989ApJ...342..807B,1995A&A...294L..41A,1996ApJ...459L..83C,2008arXiv0812.4457P,2009PhRvD..80f3005M,2009PhRvL.103e1101Y,2010arXiv1002.1910D} or SNRs \citep[\eg][]{2003A&A...410..189B,2009PhRvL.103e1104B},
while some other spatial or solar effects might also play a 
role~\citep[\eg][]{2009PhRvL.103k1302S,2010arXiv1005.4668R}. These works
have notably shown that a very few sources may dominate the high energy 
positron flux at the Earth, opening interesting perspectives for more
accurate predictions in the near future~\citep[see][for a detailed analysis]{2010arXiv1002.1910D}. Nevertheless, these perspectives are based on the assumption
that the secondary positron flux prediction is under control.

Secondary positrons originate from nuclear interactions of cosmic-ray nuclei
with the interstellar gas, and have been investigated into details 
by~\citet{1998ApJ...493..694M}, and more recently by 
\citet{2009A&A...501..821D}, who provided more insights on the theoretical 
uncertainties. \citet{2010arXiv1002.1910D} have improved the predictions of the 
latter by including the Klein-Nishina corrections to the energy loss treatment. 
All these predictions rely on cosmic-ray nuclei spectra constrained from rather 
low energy data ($\lesssim 100$ GeV), which are mostly power laws. Nevertheless,
two balloon experiments, ATIC2~\citep{2009BRASP..73..564P} and CREAM~\citep{2010ApJ...714L..89A}, have recently reported on a clear and almost concordant 
hardening in these spectra around a few TeV, with a very good statistics.
Since high energy stable nuclei have quite long range propagation
\citep[\eg][]{2003A&A...402..971T}, it is likely that this spectral inflection
is not merely local and pertains over a few kpc scale around the Earth. 
Therefore, a consistent prediction of the secondary positron flux should
take it into account.

In this paper, we study into detail the impact of these new cosmic ray 
measurements on the secondary positron flux predictions. We provide the reader 
with user-friendly fitting formulae which summarize our results, and which
can be used to constrain any extra source of cosmic-ray positrons.

\section{From the hardening of cosmic-ray nuclei spectra to the hardening of 
  the secondary positron spectrum}
\label{sec:flux}

\subsection{Generalities}
\label{subsec:gen}

Predictions of secondary positrons are usually valid in the frame of a
propagation model, which specifies the way the propagation equation is solved.
For more insights on cosmic ray propagation, we refer the reader to~\eg\
\citet{1964ocr..book.....G,berezinsky_book_90,1994hea..book.....L}.
Here, we adopt the formalism described in~\citet{2010arXiv1002.1910D}, where
convection and diffusive reacceleration are neglected, which is known to
be a good approximation in the GeV-TeV energy energy range 
\citep{2009A&A...501..821D}, and where fully relativistic energy losses are 
considered. Aside from energy losses, our propagation ingredients are therefore 
the diffusion coefficient $K(E)$, that we take homogeneous, and the 
half-thickness $L$ of the diffusion slab. We use sets of propagation 
parameters consistent with the analysis of the secondary-to-primary nuclei 
ratios performed by~\citet{2001ApJ...555..585M}, and further used 
in~\citet{2004PhRvD..69f3501D} to define the {\em minimal}, {\em median} and 
{\em maximal} sets widely used in the literature. We note that the 
\propmed~model, which was the best-fit model found 
in~\citet{2001ApJ...555..585M}, has properties very similar to the best-fit 
model derived more recently in~\cite{2010A&A...516A..66P} from a Markov chain 
Monte Carlo analysis. Details on the effects of these parameters on the 
electron and positron propagation are given in~\citet{2010arXiv1002.1910D}. 
For the energy losses, we adopt the model denoted M1 in this reference.

Our whole propagation framework can be encoded in the form of a Green function
\greenff{E,\vec{x}_\odot}{E_s,\vec{x}_s}~that characterizes the probability of a
positron injected at any coordinate $\vec{x}_s$ with energy $E_s$ to reach
an observer on Earth with energy $E\leq E_s$. This allows us to write the 
positron flux as the following convolution:
\ben
\phi(E) = \int_{\rm slab} d^3\vec{x}_s \int dE_s\,
\greenff{E,\vec{x}_\odot}{E_s,\vec{x}_s}\,{\cal Q}(E_s,\vec{x}_s)\;,
\label{eq:phi_sec}
\een
where ${\cal Q}$ is the source term that we are going to determine in the 
following. For secondaries, it formally reads~\citep[\eg][]{2009A&A...501..821D}
\ben
{\cal Q}(E,\vec{x}) = 4\pi \sum_{i,j}\int dE_k \, \phi_i(E_k) \,
\frac{d\sigma_{ij}}{dE}(E_k\to E)\,n_j(\vec{x})\;,
\label{eq:q_sec}
\een
where $\phi_i$ is the flux of a cosmic-ray species of index $i$, $n_j$ is the 
interstellar density of a gas species of index $j$, and $d\sigma_{ij}$ is the
inclusive nuclear cross section associated with the production of a positron
of energy $E$ from an ion of kinetic energy $E_k$. For the cosmic-ray nuclei 
and interstellar gas, we can safely consider the dominant species only, \ie~the 
protons and $\alpha$ ions on the one hand, and the hydrogen (90\%) and helium 
(10\%) gas on the other hand. As in \citet{2009A&A...501..821D} we assume an 
overall gas density of $n_0 = 1\,\percc$ homogeneously distributed inside an 
infinite flat disk of half-thickness $h=100$ pc, such that $n_{\rm tot}(\vec{x})
= 2\,h\,n_0\,\delta(z)$, where $z$ is the coordinate perpendicular to the 
Galactic plane. These values are justified either by measurements of the 
interstellar medium~\citep{2001RvMP...73.1031F} and the fact that high energy 
positrons have very short range propagation due to efficient energy losses
--- large scale fluctuations of the gas density have almost no effect on the 
local high energy positron flux.

\begin{figure*}
  \centering
  \includegraphics[width=\columnwidth]{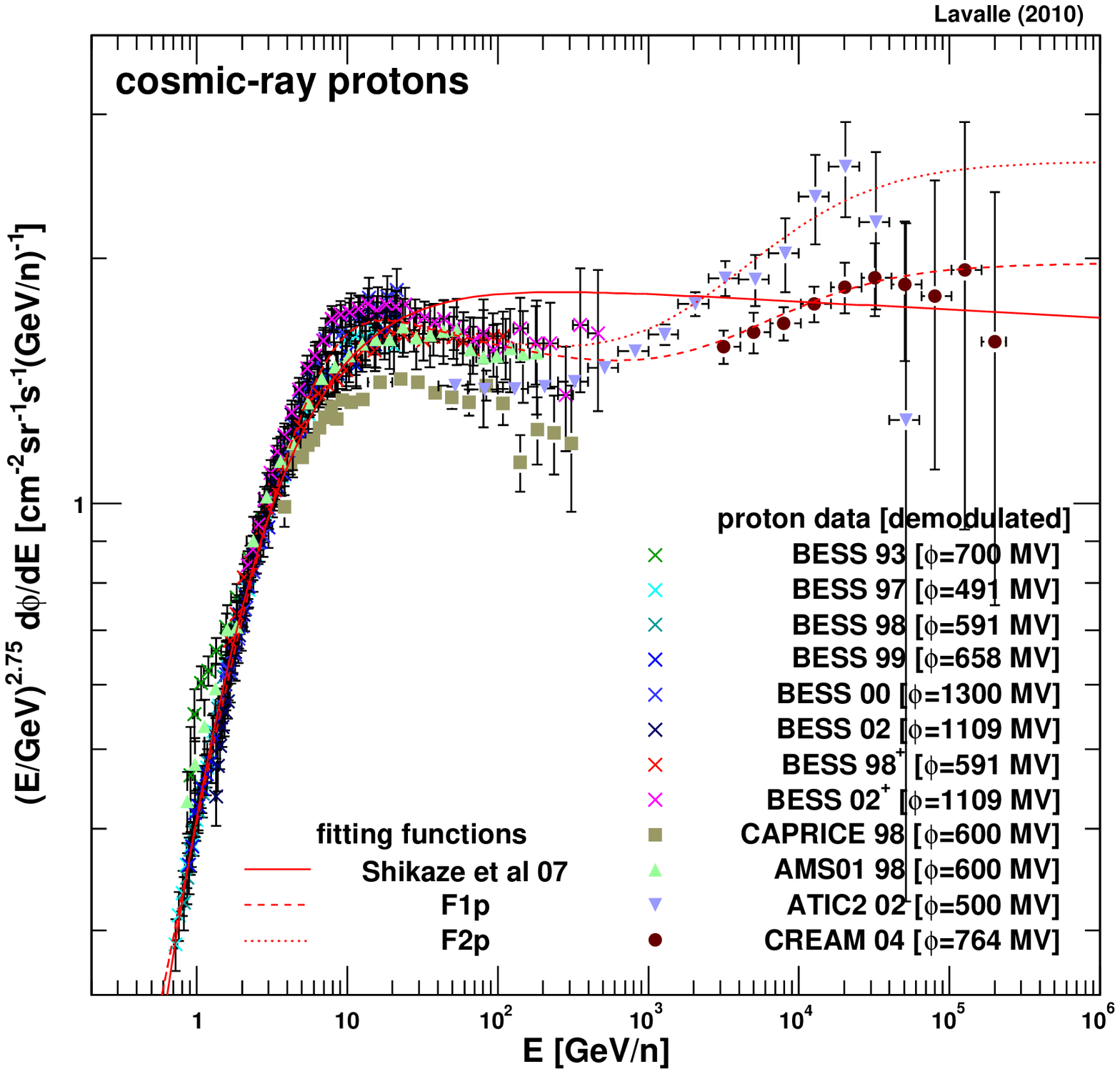}
  \includegraphics[width=\columnwidth]{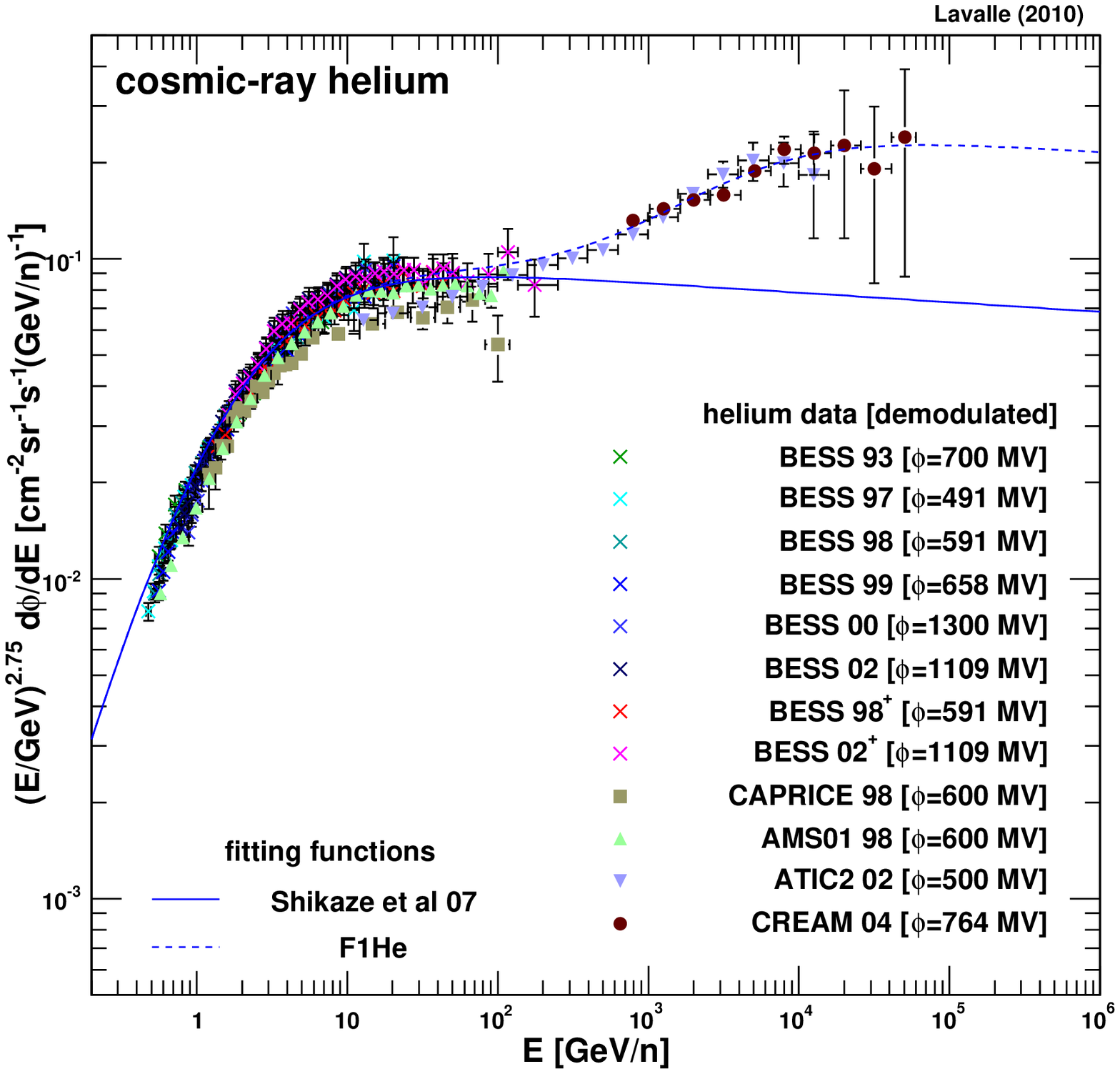}
  \caption{Left panel: cosmic-ray proton data. Right panel: cosmic-ray $\alpha$
    data.}
  \label{fig:cr_spectra}
\end{figure*}

\subsection{The incident cosmic ray flux}

In \citet{2009A&A...501..821D}, we solved \citeeqss{eq:phi_sec}{eq:q_sec} 
assuming the proton and $\alpha$ fluxes as fitted in~\citet{2007APh....28..154S}
to the low energy BESS data~\citep{2000ApJ...545.1135S,2002ApJ...564..244W,2004PhLB..594...35H}. These fits are recalled hereafter:
\ben
\phi_p^{\rm bess}(E_k) &=& A\,\beta^{a_1}\,
\left[ \frac{\cal R}{1\,{\rm GV}}\right]^{-a_2}\nn\\
\phi_{\alpha}^{\rm bess}(E_{k}/{\rm n}) &=& B\,
\beta^{b_1}\,\left[ \frac{\cal R}{1\,{\rm GV}}\right]^{-b_2}\;,
\label{eq:fit_bess}
\een
with $(A,a_1,a_2)=(1.94~\dphiusr,0.7,2.76)$ and $(B,b_1,b_2)= 
(0.71~\dphiusr,0.5,2.78)$, \ie~single power laws of indices -$a_2$ and -$b_2$,
respectively, at high energy.

These functions are displayed (solid curves) in~\citefig{fig:cr_spectra} 
together with the 
BESS~\citep{2000ApJ...545.1135S,2002ApJ...564..244W,2004PhLB..594...35H,2007APh....28..154S}, 
CAPRICE~\citep{1999ApJ...518..457B,2003APh....19..583B}, 
AMS01~\citep{2002PhR...366..331A},
ATIC2~\citep{2006astro.ph.12377P,2009BRASP..73..564P},
and CREAM~\citep{2010ApJ...714L..89A} data --- note that these data have been
corrected for solar modulation effects (demodulated) by means of the Force Field
approximation~\citep{1968ApJ...154.1011G,1971JGR....76..221F}, with Fisk 
potentials made explicit in the plot. It is clear that though these
parameterizations provide reasonably good fits to the low energy data, they 
completely fail above a few tens of GeV. In particular, we can remark that
the proton data (left panel) are overshot between $\sim 10$ GeV and a few TeV, 
while the helium data (right panel) are underpredicted above $\sim 100$ GeV/n. 
Moreover, though CREAM and ATIC2 seem to agree in their measurements of the 
helium flux, there is an unequivocal discrepancy in the proton flux.

Since the functions of~\citeeq{eq:fit_bess} were used 
in~\citet{2009A&A...501..821D} and~\citet{2010arXiv1002.1910D} to improve the
predictions of the secondary positron flux, it is worth revisiting these
predictions again in light of the new cosmic ray data. For the inclusive nuclear
cross sections, we still use the numerical approach presented 
in~\citet{2006ApJ...647..692K} for the proton-proton collision, with the 
correction prescriptions of~\citet{2007NIMPB.254..187N} for nucleus-nucleus 
interactions. The change in the derivation of the positron source term defined 
in~\citeeq{eq:q_sec} will therefore only come from the updated fits of the 
proton and helium fluxes.

To illustrate the difference in the positron source term arising from 
considering either the CREAM or the ATIC2 proton data, we consider two different
modelings defined by the preference we put in one or the other experiment. The 
proton flux parameterization will be the only change between these two cases, 
since the helium data of the two experiments are consistent and can be 
simultaneously fitted by the same function.

In the following, we denote F1p and F2p the parameterizations associated
with the proton data, the former providing a good fit to the CREAM data,
and the latter providing a good fit to the ATIC2 data. We also define a function
F1He that provides a good fit to both CREAM and ATIC2 helium data. Functions 
F1p, F2p and F1He are given in \citesec{subsec:cr_fits}, and are displayed 
in~\citefig{fig:cr_spectra}, in the left panel for the two first, and in the 
right panel for the last one. The functions of~\citeeq{eq:fit_bess} will be 
referred to as {\em low energy fit}, or {\em reference fit}. In contrast,
the combination of F1p and F1He will be referred to as {\em CREAM fit},
while F2p and FHe as {\em ATIC2 fit}.

\subsection{Impact of the cosmic ray hardening on the secondary positron 
  source term}

\begin{figure*}
  \centering
  \includegraphics[width=\columnwidth]{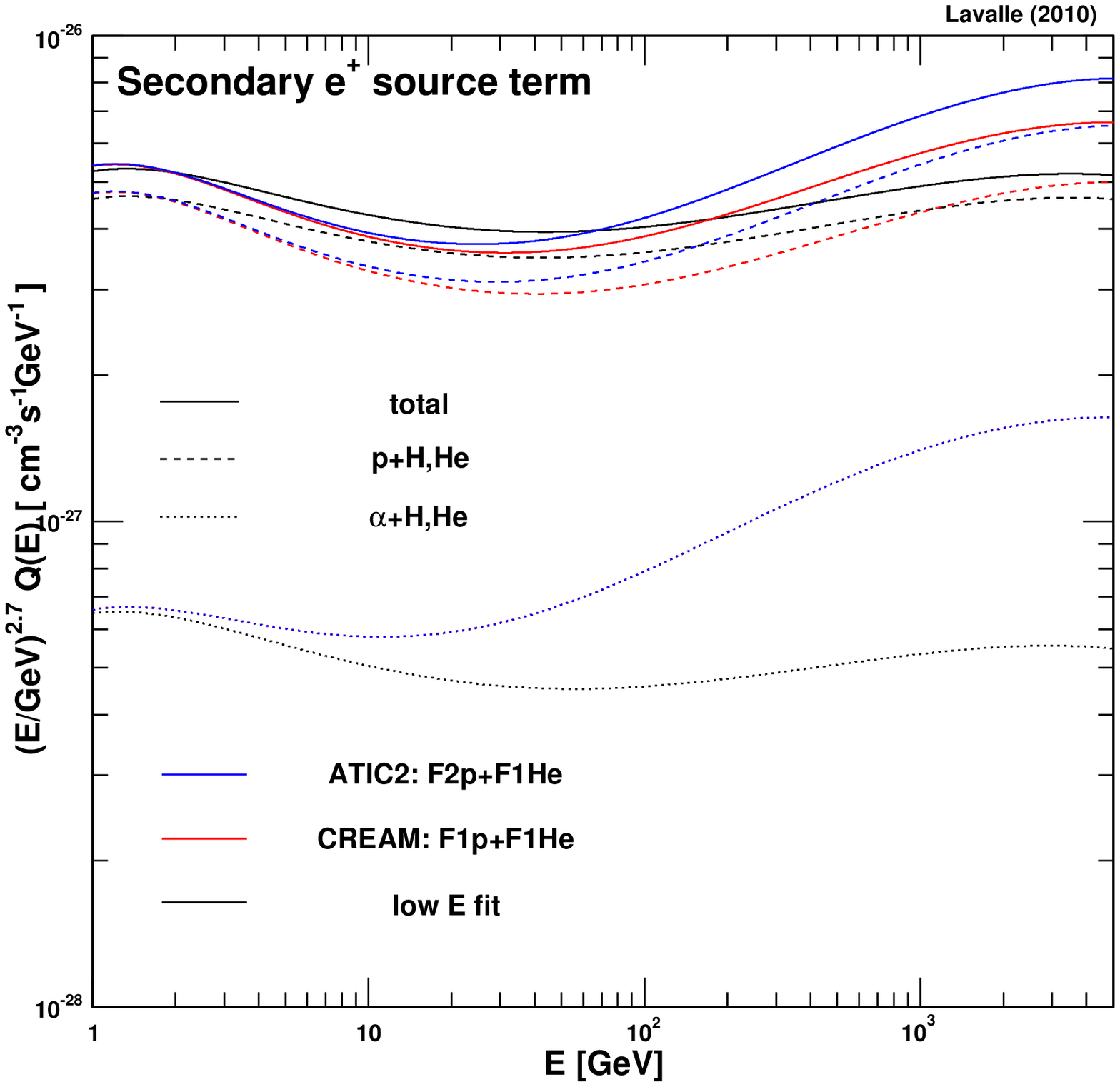}
  \includegraphics[width=\columnwidth]{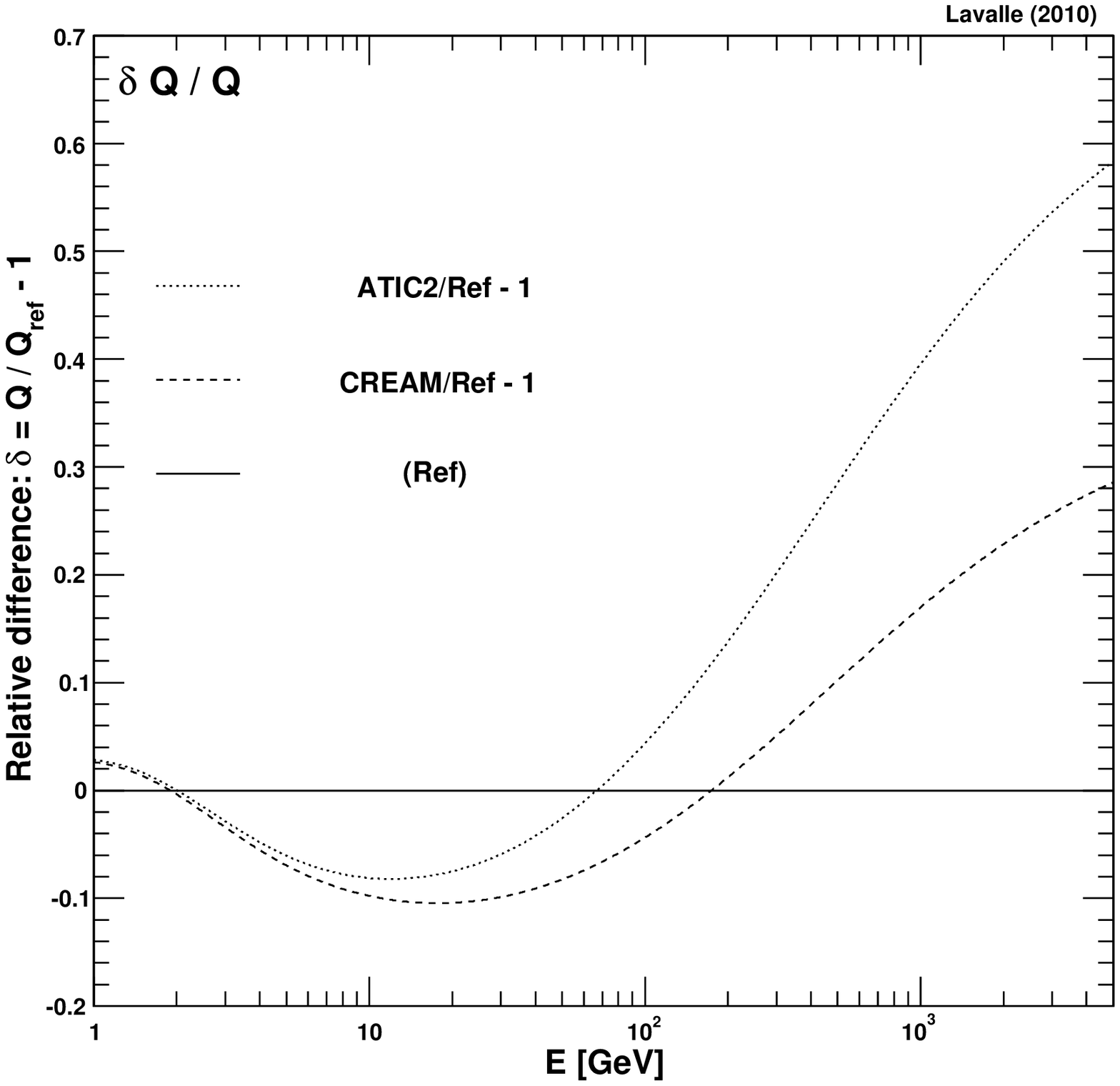}
  \caption{Left panel: Secondary positron source term for three different
    fit-based assumptions for the incident cosmic ray spectra; the contributions
    due to cosmic-ray protons (dashed curves) and $\alpha$ particles (dotted 
    curves) are shown explicitly aside from the overall contributions (solid
    curves). Right panel: relative difference between the obtained overall 
    source terms.}
  \label{fig:source}
\end{figure*}

In the left panel of~\citefig{fig:source}, we derive the secondary positron 
source term defined in~\citeeq{eq:q_sec} and associated with the low energy fit 
(black curve), the CREAM fit (red curve), and the ATIC2 fit (blue curve); note 
that all of them run close to a power law in energy of index -2.7. The 
contribution coming from the cosmic-ray proton is shown explicitly (dashed 
curves) for each modeling, as well as the $\alpha$ contribution (dotted curves).
We see that the relative increase in the $\alpha$ contribution, from the low 
energy fit to F1He, is very large, reaching a factor of $\sim 3$ around a few 
TeV. Nevertheless, it remains subdominant with respect to the proton 
contribution. The difference in the proton flux modeling translates almost 
linearly into the positron source term, leading to a relative decrease with 
respect to the low energy fit below 100 GeV (1 TeV), and a relative increase 
above, for the ATIC2 (CREAM, respectively) fit. This comes from the fact that 
the inclusive cross section featuring in~\citeeq{eq:q_sec} scales like $\sim 
1/E_k$ \citep{2009A&A...501..821D}, straightforwardly leading to 
${\cal Q}\overset{\sim}{\propto} \phi_i(E_k)$; this explains why all 
contributions almost scale like $E^{-2.7}$. When summing up the contributions
coming from proton and $\alpha$ interactions, we see that the relative decrease
apparent in the proton-only case is less prominent due to the positive 
yield, though modest, of the $\alpha$ interactions: the net effect is a very 
slight decrease below 100 GeV, and a larger increase above. This is illustrated
in more detail in the right panel of \citefig{fig:source}, where we plot the 
relative difference of the CREAM (dashed curve) and ATIC2 (dotted curve) cosmic
ray-induced positron source terms with the low energy reference case: the slight
decrease below 100 GeV makes a 10\% difference at most with the reference case, 
while the increase above reaches $\sim 30\%$ (60\%) above a few TeV for the 
CREAM (ATIC2, respectively) configuration. The impact of using these new cosmic 
ray data is therefore not negligible in terms of secondary positron production.

\subsection{Updated predictions for the secondary positron flux}
\label{subsec:pos_flux}

Because of propagation, guessing the effect of the new cosmic ray modelings
on the secondary positron flux prediction might not look, {\it a priori}, as 
straightforward as guessing their effect on the positron source term. This is 
formally due to the non-trivial dependence of the positron Green function on 
energy. We refer the reader to \citet{2010arXiv1002.1910D}, in particular to 
their Sect. 2.3, for more insights about this dependence. Nevertheless, since 
here we deal with a source term which is homogeneously distributed inside a thin
disk of half-thickness $h$, and with an energy dependence close to a 
power law, we can assume that 
${\cal Q}(E,\vec{x} \approx 2\,h\,{\cal Q}_0\,\delta(z)\,
(E/{\rm GeV})^{-\gamma}$. In that case, the positron flux at the Earth can be 
approximated as \citep{2010arXiv1002.1910D}:
\ben
\phi_\odot(E) \overset{\sim}{\propto} 
\frac{c\,h\,{\cal Q}_0}{\sqrt{K_0/\tau_l}} \, (E/{\rm GeV})^{-\tilde{\gamma}}\;,
\label{eq:flux_approx}
\een
where $K_0$ is the normalization of the diffusion coefficient, $\tau_l$ is
the energy loss timescale, and $\tilde{\gamma} \approx \gamma + 0.5\,(\alpha+
\delta-1)$ is the predicted flux spectral index which depends on the injected 
index $\gamma$ and also on the diffusion coefficient index $\delta$ and
the energy loss index $\alpha$ ($\alpha = 2$ in the Thomson approximation, 
but is $< 2$ when Klein-Nishina corrections become sizable).

The above equation is of great interest to anticipate the coming results, 
since it tells us that the ratio of two different flux predictions scales like 
the ratio of the corresponding source terms at any energy. Therefore, the 
relative differences in flux predictions should be very close to the relative 
differences in the corresponding source terms, the latter being already plotted 
in the right panel of \citefig{fig:source}. This is actually verified in the 
right panel of~\citefig{fig:flux}, which can hardly be differentiated from the 
one previously mentioned, illustrating how efficient the approximation given in
\citeeq{eq:flux_approx} is in this context. In the left panel 
of~\citefig{fig:flux}, we show the corresponding predictions for the 
secondary positron flux, which are derived with the \propmed~propagation 
parameters and with model M1 for the energy losses --- these parameters 
can be found in~\citet{2010arXiv1002.1910D}. From these plots, we readily 
conclude that the secondary positron flux prediction is affected over the whole 
GeV-TeV energy range, decreasing by $\sim 10$\% below $\sim 100$ GeV, and 
increasing by more than 30\% at TeV energies, up to 60\% in the case of the 
ATIC2 cosmic ray fit. The spectral index is therefore increased accordingly 
by almost 2 digits, from $\sim$-3.5 to $\sim$-3.3 in the \propmed~case 
employed here.

\begin{figure*}
  \centering
  \includegraphics[width=\columnwidth]{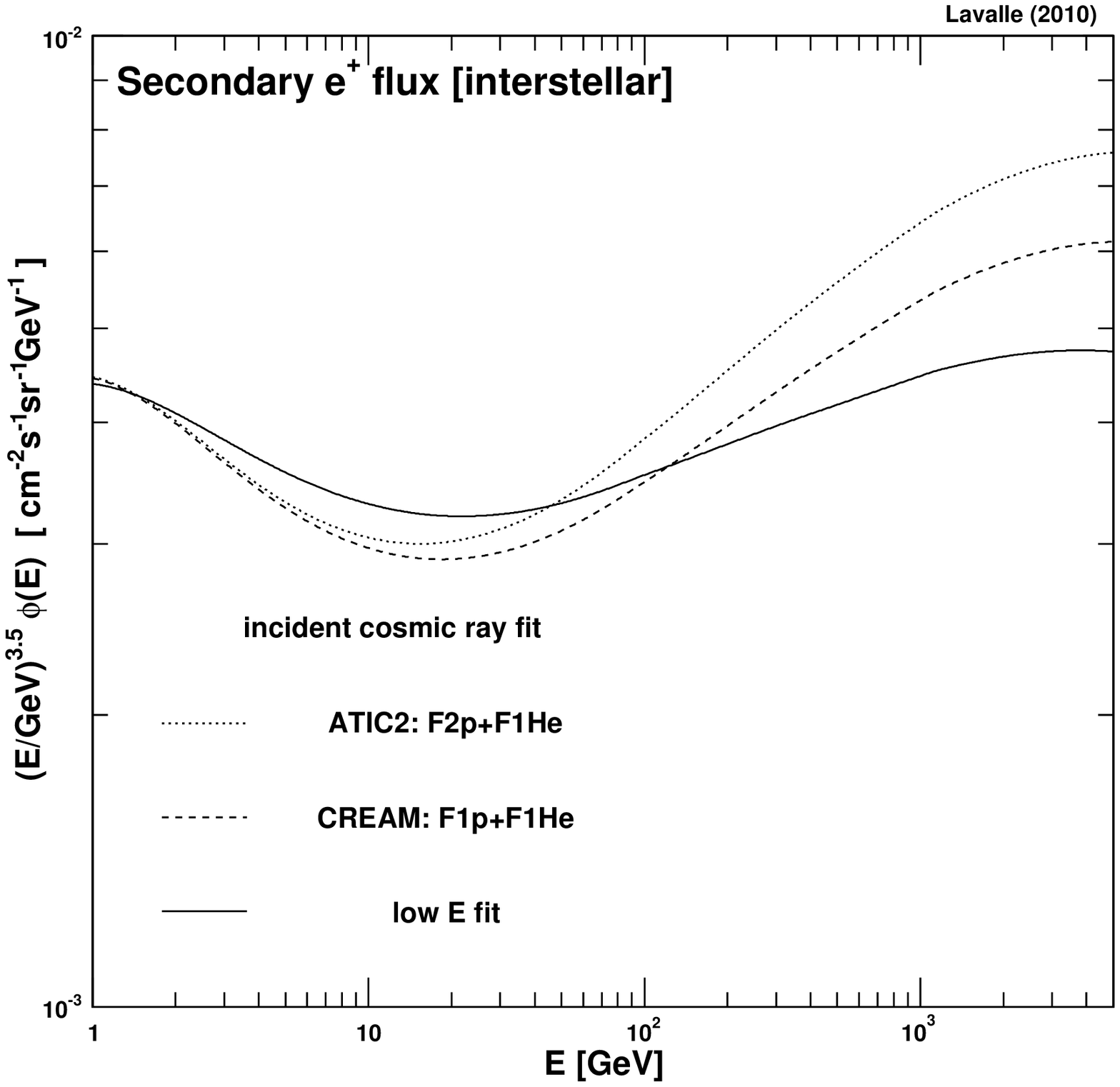}
  \includegraphics[width=\columnwidth]{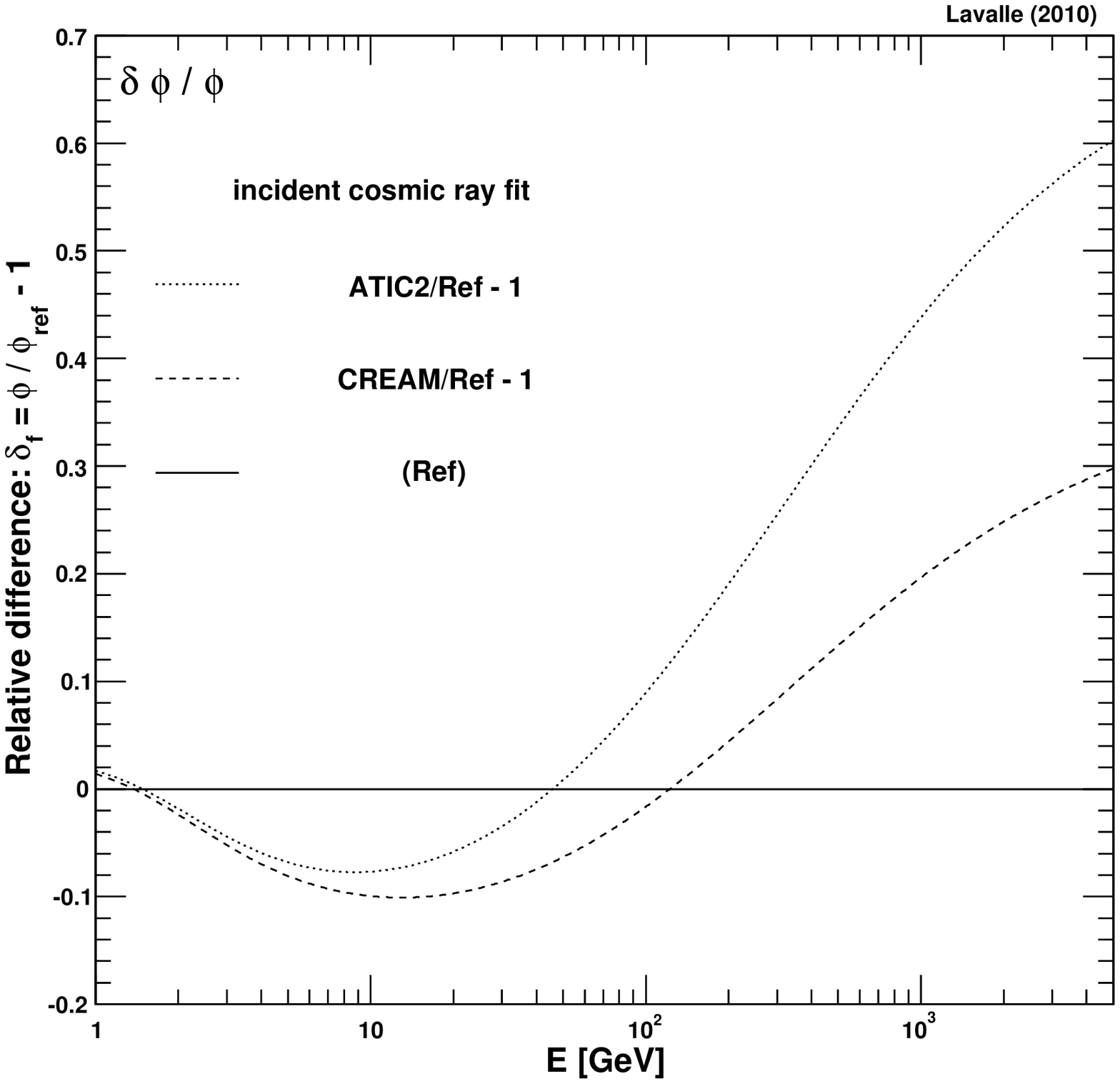}
  \caption{Left panel: Secondary positron flux predictions associated with 
    the 3 different fit-based assumptions considered for the incident cosmic 
    ray spectra; the~\propmed propagation setup has been used. Right panel: 
    relative difference between the predictions shown in the left panel.}
  \label{fig:flux}
\end{figure*}

For the sake of completeness, it is interesting to derive the theoretical
uncertainty bands associated with these novel predictions, which come from the 
uncertainties in the propagation parameters. To this aim, we proceed as 
in~\citet{2009A&A...501..821D} by bracketing the \propmed~with the 
\propmin~and~\propmax~propagation configurations. We report our results for the
positron flux (fraction) in the top (bottom,respectively) panels of 
\citefig{fig:flux_frac} --- a solar modulation is applied, using the Force 
Field approximation \citep{1968ApJ...154.1011G,1971JGR....76..221F} with a Fisk 
potential of 600 MV, and ignoring any charge dependence effect potentially 
important below a few GeV~\citep[\eg][]{2006SSRv..127..117H}. The positron flux 
data are
taken from~~\citet{2000ApJ...532..653B,2001ApJ...559..296D,2002PhR...366..331A},
while the positron fraction data are from~\citet{1997ApJ...482L.191B,2004PhRvL..93x1102B,2007PhLB..646..145A,2010APh....34....1A}. To calculate the positron
fraction $f=\phi_{e^+}/(\phi_{e^+}+\phi_{e^-})$, we determined the denominator
above 10 GeV from the Fermi-LAT 
data~\citep{2009PhRvL.102r1101A,2010arXiv1008.3999F}, and used the AMS01 data 
\citep{2002PhR...366..331A}~to constrain the electron flux at lower energy.
In all panels, we also display the result obtained 
by~\citet{1998ApJ...493..694M}, as fitted by~\citet{1998PhRvD..59b3511B},
since it is still very often used in the literature as a reference. Note that 
the predictions obtained from the low energy cosmic ray fit are obviously 
identical to the ones derived in~\citet{2010arXiv1002.1910D}.

From the predictions based on the low energy cosmic ray fit to the ones 
based on the ATIC2 fit, the secondary positron spectrum is hardened, which 
translates into a slightly flatter positron fraction. Of course, we did not 
expect the present study to be relevant to the discussion on the rise of the 
positron fraction itself, since the enhancement in the secondary positron
flux was already known to be by far too small from the incident cosmic-ray 
nuclei data. It is instead very useful so as to constrain any extra 
source of positrons, like pulsars or dark matter, with the data. Indeed, in 
that case, one 
needs to add a secondary contribution to the primary one in a consistent manner
before comparing the sum to the data. From the top panels of~\citefig{fig:flux_frac}, we note incidentally 
that the \propmin~propagation setup already leads to a conflict with the data 
because of its too small value of $K_0$ [see~\citeeq{eq:flux_approx}] 
(associated with a 
small value of $L$, $K_0/L$ being roughly fixed by secondary-to-primary 
cosmic-ray nuclei ratios). This configuration is in any case obsolete, since it 
is no longer supported by recent secondary-to-primary analyses
\citep[\eg][]{2010A&A...516A..66P}, nor by the Fermi-LAT diffuse gamma-ray data
~\citep[\eg][]{2010arXiv1011.0816F} (see also a dedicated discussion
in~\citealt{2010PhRvD..82h1302L}). Nevertheless, it can still be thought
of as an extreme configuration and used for illustration purposes.

We provide user-friendly empirical fitting functions associated with all these 
propagation models in~\citesec{subsec:pos_fits}.

\begin{figure*}
  \centering
  \includegraphics[width=0.66\columnwidth]{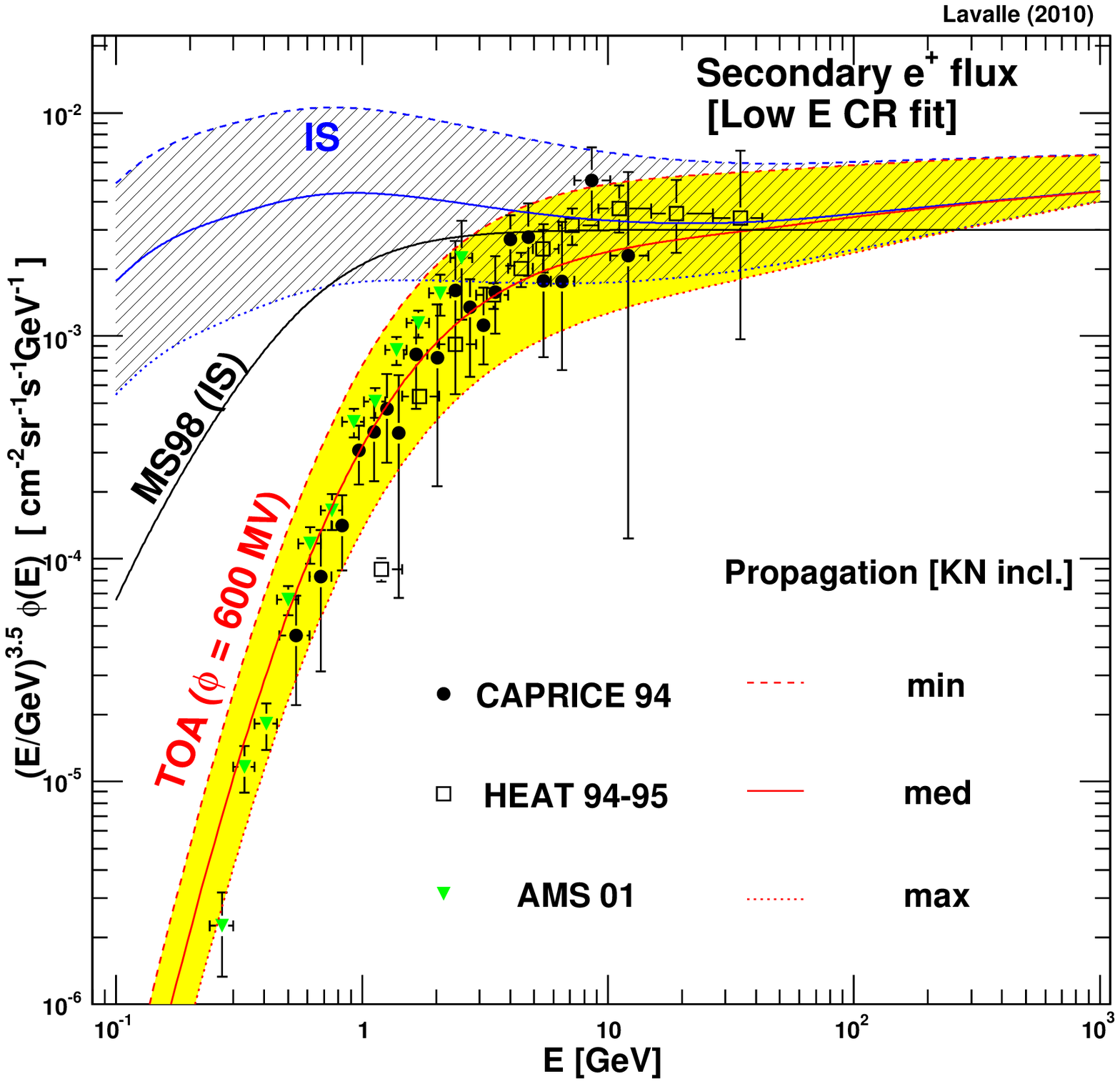}
  \includegraphics[width=0.66\columnwidth]{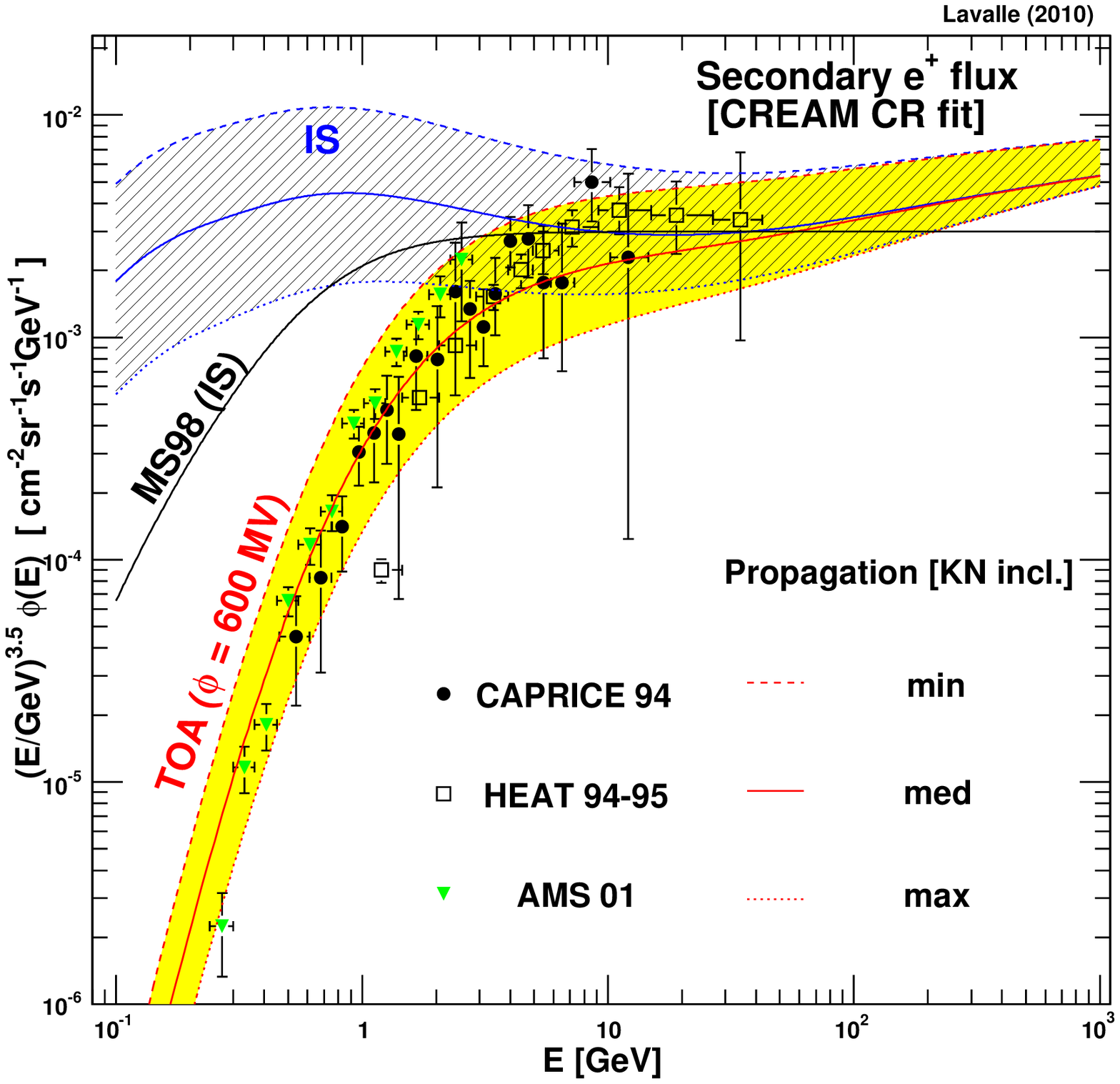}
  \includegraphics[width=0.66\columnwidth]{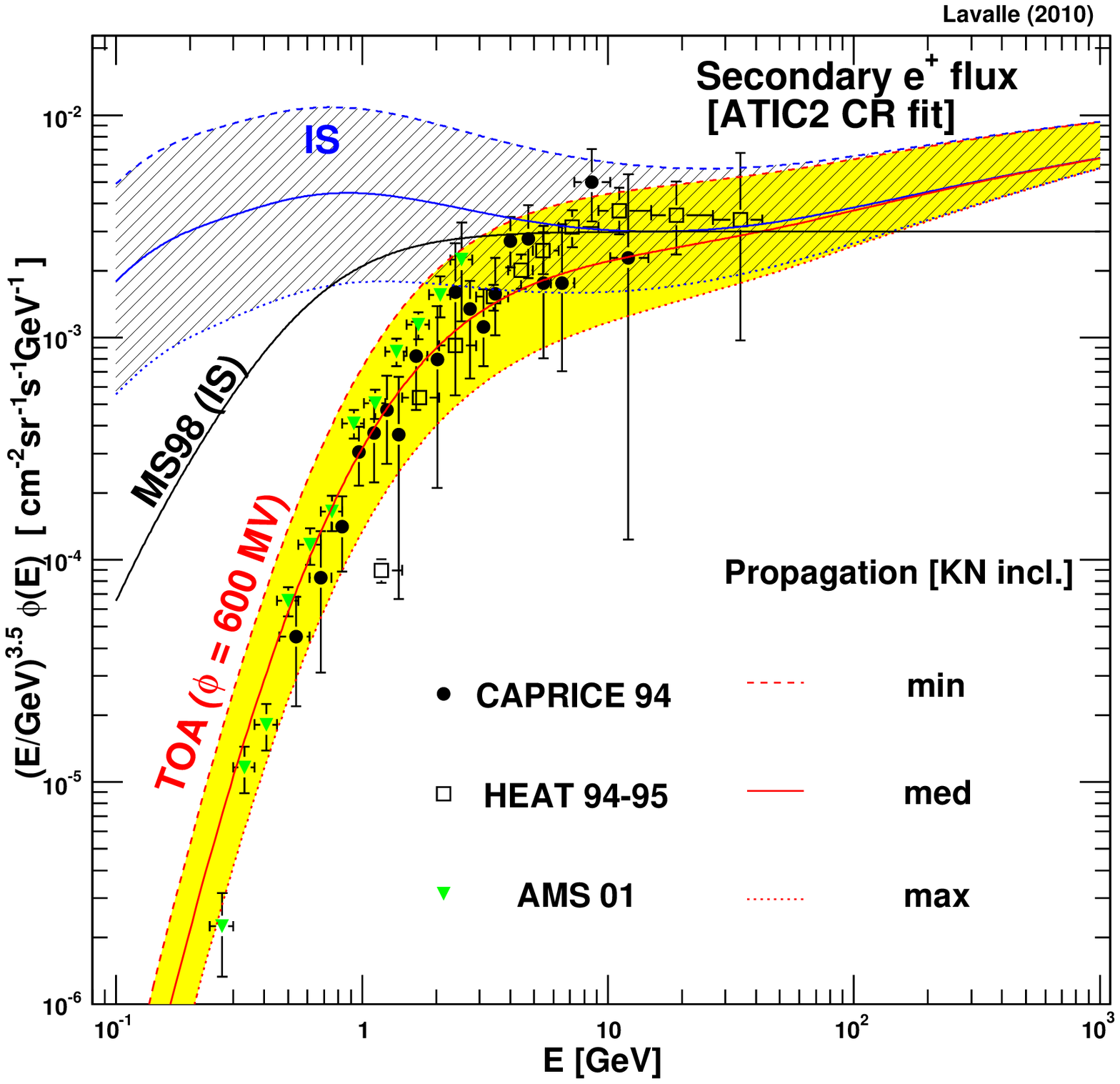}
  \includegraphics[width=0.66\columnwidth]{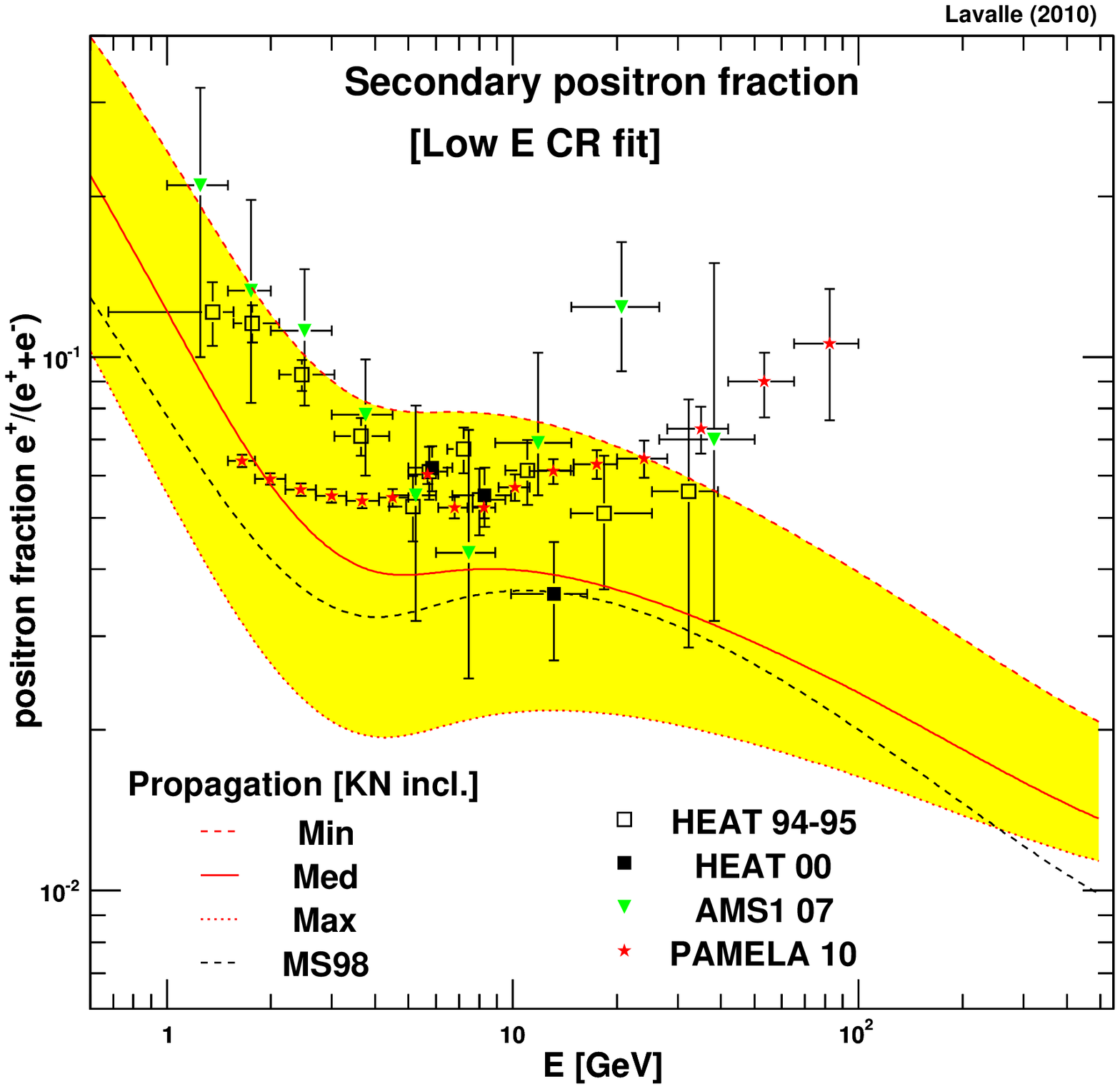}
  \includegraphics[width=0.66\columnwidth]{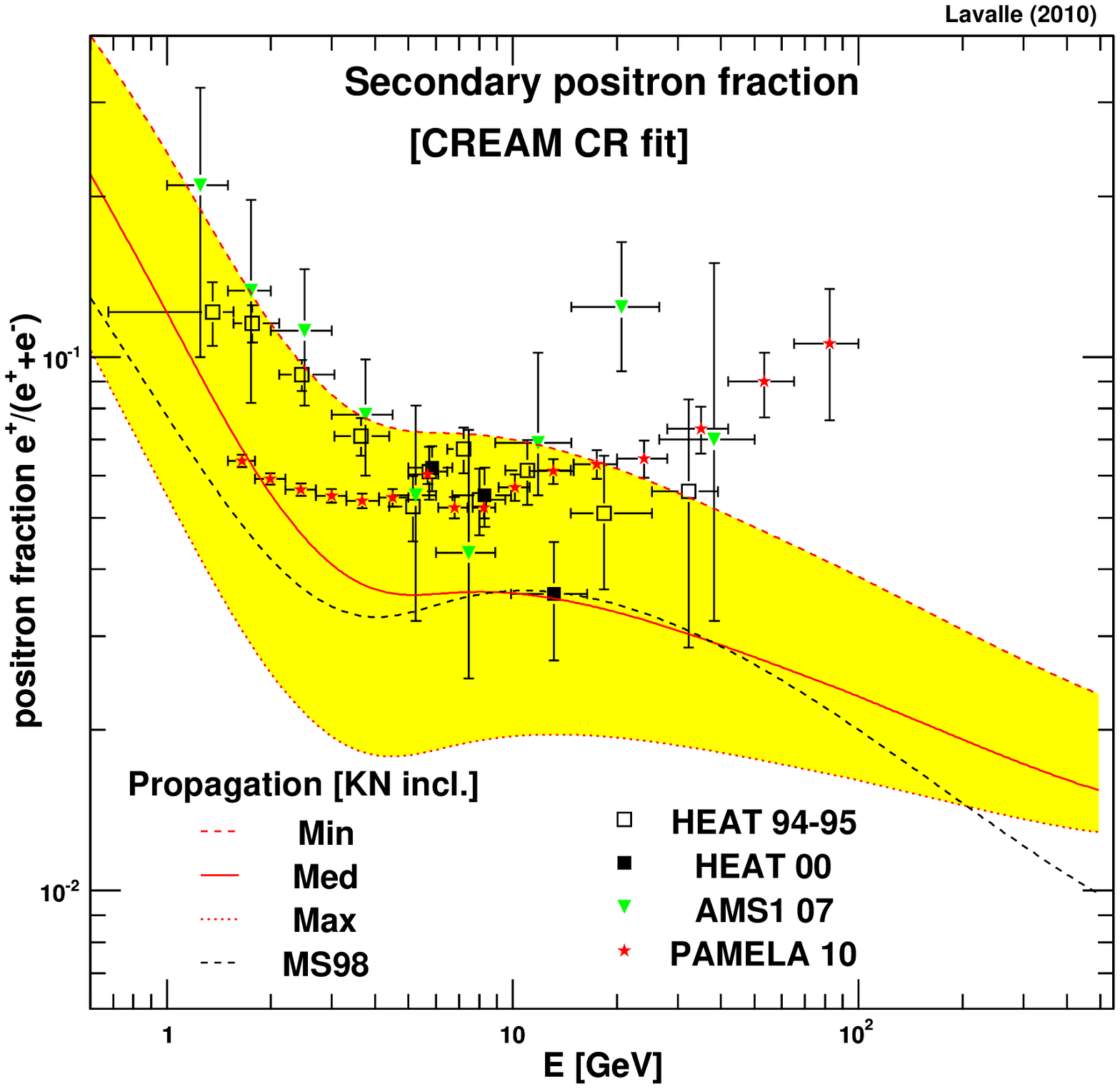}
  \includegraphics[width=0.66\columnwidth]{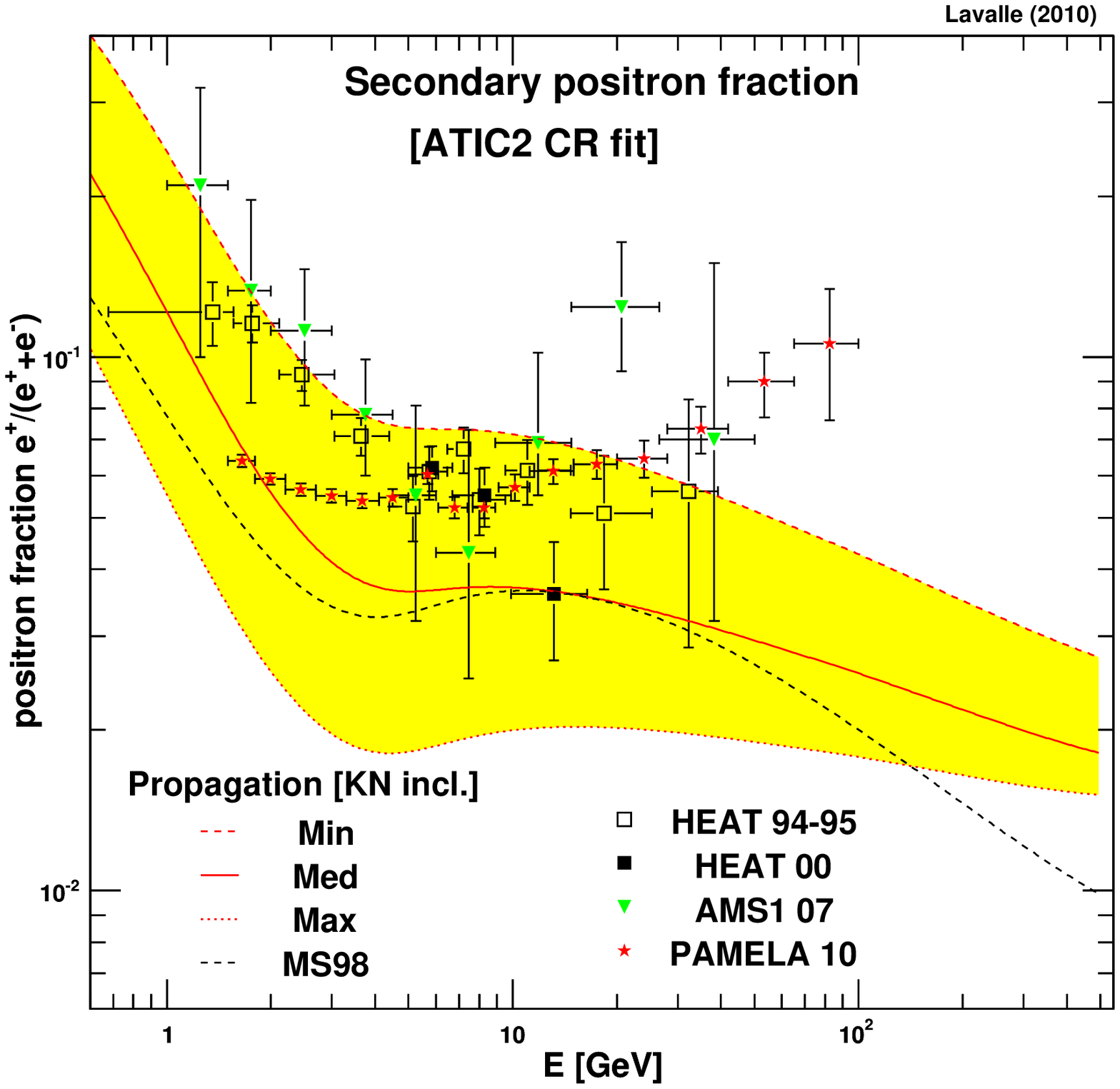}
  \caption{Top panels: secondary positron flux predictions, both interstellar 
    (IS) and on top of atmosphere (TOA), assuming a Fisk potential of 600 MV
    for the solar modulation. Bottom panels: associated positron fractions. 
    These predictions correspond to source terms calculated from either the low 
    energy cosmic ray fit (left panels), the CREAM cosmic ray fit (middle 
    panels), or the ATIC2 cosmic ray fit (right panels).}
  \label{fig:flux_frac}
\end{figure*}

\section{User-friendly fitting formulae}
\label{sec:fits}

In this section, we provide the reader with the parametric functions (i) that we
used to fit the cosmic ray data and (ii) also the ones which fit our prediction 
results for the secondary positron flux.

\subsection{Cosmic-ray proton and helium interstellar fluxes}
\label{subsec:cr_fits}

The proton data can be accommodated by two different empirical functions, 
depending on whether one favors the CREAM (F1p) or ATIC2 (F2p) data at high 
energy. These functions are both characterized by the following 
parameterization:
\ben
{\rm Fp}(E_k) &=& \phi^0_p\,
\left[1-e^{-\left(\frac{E_k}{E_{p1}}\right)^{p_1}} \right]\,
\left[\frac{E_k}{10\,\gev/{\rm n}}\right]^{-\gamma_p}\nn\\
&\times& \left[1+\frac{E_k}{E_{p2}}\right]^{p_2}\,
\left[1+\frac{E_k}{E_{p3}}\right]^{p_3}\;.
\label{eq:fit_p}
\een
One recognizes a standard power law of main index $\gamma_p$ associated with an 
exponential attenuation factor active at low energy, which clearly improves the 
low energy fit with respect to~\citeeq{eq:fit_bess}, and a double spectral
correction of indices $p_2$ and $p_3$ operating above kinetic energies 
$E_{p2}$ and $E_{p3}$, respectively. Note that function Fp behaves 
asymptotically as a power law of index $\gamma_\infty = -\gamma+p_2+p_3$.

In contrast, a single function (F1He) is enough in the case of helium ions,
since both sets of high energy data are in agreement. In that case,
a slight correction to~\citeeq{eq:fit_bess} is enough, so that
\ben
{\rm FHe}(E_k/{\rm n}) = \phi_{\alpha}^{\rm bess}(E_{k}/{\rm n})\,
\left[1+\frac{\cal R}{{\cal R}_{p2}}\right]^{p_2}\,
\left[1+\frac{\cal R}{{\cal R}_{p3}}\right]^{p_3}\;,
\label{eq:fit_he}
\een
where $\phi_{\alpha}^{\rm bess}$ is given in~\citeeq{eq:fit_bess}, and 
${\cal R}$ is the rigidity. There is also a double spectral correction
as in the case of protons. As function Fp, function FHe behaves asymptotically
as a power law of index $\gamma_\infty = -b_2+p_2+p_3$, where $b_2$ is defined
in \citeeq{eq:fit_bess}.

Given the spread in the available data, we did not performed a $\chi^2$ 
selection of the parameters. Indeed, this would require an {\em a priori}
or expert selection of the data, which is beyond the scope of this paper.
The values of the parameters used for functions F1p, F2p and F1He are listed 
in~\citetab{tab:fit_cr}.

\begin{table*}
\centering
\begin{tabular}{ccccccccc}
\hline
&$\phi^0$& $\alpha_p$ & $E_{p1}$ & $p_1$ & $E_{p2}$ & $p_2$ & $E_{p3}$ & $p_3$\\
&$[\dphiusr]$ &&[GeV]&&[TeV]&&[TeV]\\
\hline
F1p &  $3.09\times 10^{-3}$ & 2.8 & 4 & 1.05 & 2.5 & 0.34 & 10 & -0.29 \\
\hline
F2p &  $3.09\times 10^{-3}$ & 2.8 & 4 & 1.05 & 1.5 & 0.4  & 10 & -0.35 \\
\hline
F1He & - & - & - & - & 1 & 0.5 & 10 & -0.5 \\
\hline
\end{tabular}
\caption{Parameters used to fit the cosmic-ray proton (F1p, F2p) and helium 
  (F1He) data, according to~\citeeqss{eq:fit_p}{eq:fit_he}. For function F1He,
  any parameter $E_{p}$ (in GeV) in this table corresponds to ${\cal R}_p$ 
  (in GV) in the associated equation. }
\label{tab:fit_cr}
\end{table*}

\subsection{Secondary positron flux predictions}
\label{subsec:pos_fits}

Here, we define an empirical function that provides a very good fit to
the {\em interstellar} secondary positron flux predictions presented 
in~\citesec{subsec:pos_flux}:
\ben
\phi_{\rm fit}(E) = \exp\left\{ \sum_{i=0} c_i 
\left[ \ln\left( \frac{E}{\rm GeV} \right) \right]^i \right\}\;.
\label{eq:fit_pos}
\een
The parameters associated with all the configurations discussed in
\citesec{subsec:pos_flux} are given in~\citetab{tab:fit_pos}.
We emphasize that these parameters are valid only for the propagation models,
energy loss configurations, and incident cosmic ray modelings discussed
through this paper. These empirical fitting functions should not be used for
other sets of parameters. They are valid from $\sim 0.1$ GeV to $\sim$ 10 TeV.

\begin{table*}
\centering
\begin{tabular}{c|ccc|ccc|ccc}
\hline
 &     & Low $E$ CRs & &     & CREAM CRs &      &      & ATIC2 CRs & \\
%\hline
 &\propmin&\propmed &\propmax& \propmin&\propmed& \propmax &\propmin& \propmed    & \propmax \\
\hline
$c_0$ & -4.61 & -5.48 & -6.4 & -4.61 & -5.48 & -6.4 & -4.41 & -5.48 & -6.4 \\
%\hline
$c_1$ & -3.55 & -3.486 & -3.37 & -3.6 & -3.53 & -3.41 & -3.6 & -3.53 & -3.41\\
%\hline
$c_2\,(\times 10^{-2})$&-8.59&-8.34&-8.2&-9.28&-8.99&-8.83&-8.98&-8.68&-8.52\\
%\hline
$c_3\,(\times 10^{-2})$&2.21&2.16& 2.13 & 2.73 & 2.67 & 2.64&2.79&2.73 & 2.69\\
%\hline
$c_4\,(\times 10^{-3})$&-1.41&-1.38&-1.37&-1.8&-1.77&-1.754&-1.88& -1.84&-1.82\\
\hline
\end{tabular}
\caption{Parameters used to fit the interstellar secondary positron flux 
  predictions, from~\citeeq{eq:fit_pos}. }
\label{tab:fit_pos}
\end{table*}

\section{Conclusion}
\label{sec:concl}

In this paper, we have studied the impact of the spectral hardening observed in 
the cosmic-ray proton and helium fluxes by the 
ATIC2~\citep{2006astro.ph.12377P,2009BRASP..73..564P} and CREAM
\citep{2010ApJ...714L..89A} experiments, on the secondary positron flux 
prediction. To this aim, we have revisited the calculation of the secondary 
positron source term, which spatially tracks the interstellar gas, showing
that its energy distribution roughly scales like the incident cosmic ray
spectrum. Because of the discrepancy in the proton fluxes observed by 
the CREAM and ATIC2 balloons --- in contrast, both agree on the helium flux ---
we have considered two different cosmic ray modelings: one based on a fit to 
the CREAM proton data (moderate case), and another based on a fit to the ATIC2 
proton data (maximal case), both using the same helium flux parameterization.
The former (latter) case led to a 30\% (60\%, respectively) increase in the
production rate of TeV positrons.

Then, we have propagated these positrons to the Earth, using the propagation
framework described in~\citet{2010arXiv1002.1910D}, which includes a fully
relativistic treatment of the energy losses, with spatial diffusion 
parameters as constrained in~\citet{2001ApJ...555..585M}, still consistent
with the more recent analysis performed in~\citet{2010A&A...516A..66P}. We have 
notably explained why, in this context, the relative differences in the 
differential flux predictions were almost equal to the relative differences in 
the energy-dependent positron production rate, and consequently to the relative 
differences in the considered incident cosmic ray fluxes. This led us to 
establish a
robust estimate of the effect: as for the positron production rate, the 
secondary positron flux is increased by up to 30\% (60\%) at TeV energies 
when constraining the cosmic-ray nuclei fluxes from the CREAM (ATIC2, 
respectively) data. Therefore, these predictions differs from the ones 
performed in~\citet{1998ApJ...493..694M}, \citet{2009A&A...501..821D} and
\citet{2010arXiv1002.1910D}, resulting in harder secondary positron spectra. We 
have also derived an estimate of the theoretical uncertainties. These results 
are complementary to the recent analysis of~\citet{2010arXiv1010.5679D} on the 
secondary antiproton and diffuse gamma-ray fluxes.

Finally, we have provided the reader with user-friendly parametric functions
that allow to reproduce all the results derived in this paper, so that they 
can be straightforwardly exploited. For instance, one can use these functions to
constrain any extra source of primary positrons against the existing (or
forthcoming) data in a self-consistent way --- in which case one needs to sum 
up the secondary and primary contributions. We again emphasize that these 
results are valid only in the frame of the propagation models described through 
the paper, so if used, one has to make sure that the same parameters are 
employed for the primary component.

%% donato & serpico \citet{2010arXiv1010.5679D}

\section*{Acknowledgements}
We are grateful to T. Delahaye and P. Salati for interesting discussions about 
this study. This work was supported by the Spanish MICINN’s Consolider-Ingenio 
2010 Program under grant MultiDark CSD2009-00064. We also thank the support of
the MICINN under grant FPA2009-08958, the Community of Madrid under grant 
HEPHACOS S2009/ESP-1473, and the European Union under the Marie Curie-ITN 
program PITN-GA-2009-237920.

\bibliography{lavalle_bib}

\label{lastpage}

\end{document}